\documentclass[AMA,STIX2COL]{MRM}

\articletype{Pre-Print -- submitted to Magnetic Resonance in Medicine}%

\received{\today}
\revised{}
\accepted{}
\usepackage{tikz}
\usepackage{pgfplots}

\usetikzlibrary{spy,backgrounds,arrows,decorations.pathmorphing,backgrounds,positioning,fit,matrix,calc,shapes.multipart,shapes.misc,pgfplots.statistics, pgfplots.colorbrewer}

\pgfplotsset{compat=1.18,
	boxplot/estimator=R1,
	colormap={parula}{
			rgb255=(53,42,135)
			rgb255=(15,92,221)
			rgb255=(18,125,216)
			rgb255=(7,156,207)
			rgb255=(21,177,180)
			rgb255=(89,189,140)
			rgb255=(165,190,107)
			rgb255=(225,185,82)
			rgb255=(252,206,46)
			rgb255=(249,251,14)},
	colormap={mybluered}{
			rgb255(0cm)=(0,0,180)
			rgb255(1cm)=(0,180,180)
			rgb255(2cm)=(70,180,0)
			rgb255(3cm)=(180,180,0)
			rgb255(4cm)=(255,0,0)
			rgb255(5cm)=(128,0,0)},
	colormap={mybluewhitered_m10_2}{
			rgb255(0cm)=(0,0,255)
			rgb255(10cm)=(255,255,255)
			rgb255(12cm)=(255,0,0)},
	colormap={mybluewhitered_m1_2}{
			rgb255(0cm)=(0,0,255)
			rgb255(1cm)=(255,255,255)
			rgb255(3cm)=(255,0,0)},
	colormap={mybluewhitered_0_1}{
			rgb255(0cm)=(255,255,255)
			rgb255(1cm)=(255,0,0)},
	colormap={gist_earth}{
			rgb255=(0.0,0.0,0.0)
			rgb255=(1.7085000000000001,0.0,62.424)
			rgb255=(3.3914999999999997,0.0,89.8365)
			rgb255=(5.1000000000000005,1.4789999999999999,113.985)
			rgb255=(6.8085,7.343999999999999,116.4075)
			rgb255=(8.4915,13.209,116.8665)
			rgb255=(10.200000000000001,19.074,117.3255)
			rgb255=(11.9085,24.939,117.7845)
			rgb255=(13.5915,30.804000000000002,118.2435)
			rgb255=(15.299999999999999,36.669000000000004,118.7025)
			rgb255=(16.983,42.534,119.1615)
			rgb255=(18.6915,48.3735,119.6205)
			rgb255=(20.400000000000002,53.677499999999995,120.10499999999999)
			rgb255=(22.083,58.981500000000004,120.564)
			rgb255=(23.7915,64.2855,121.02300000000001)
			rgb255=(25.5,69.58949999999999,121.482)
			rgb255=(27.183,74.8935,121.941)
			rgb255=(28.8915,79.917,122.39999999999999)
			rgb255=(30.599999999999998,84.6855,122.859)
			rgb255=(32.282999999999994,89.454,123.3435)
			rgb255=(33.9915,94.2225,123.8025)
			rgb255=(35.7,98.838,124.2615)
			rgb255=(37.383,102.86699999999999,124.7205)
			rgb255=(39.091499999999996,106.896,125.1795)
			rgb255=(40.774499999999996,110.925,125.63850000000001)
			rgb255=(42.483,114.954,126.0975)
			rgb255=(44.191500000000005,118.983,126.5565)
			rgb255=(45.8745,123.012,127.041)
			rgb255=(47.583,127.041,127.5)
			rgb255=(48.909,129.13199999999998,125.23049999999999)
			rgb255=(50.1075,130.5855,122.1195)
			rgb255=(51.306,132.0645,118.983)
			rgb255=(52.5045,133.518,115.872)
			rgb255=(53.7285,134.99699999999999,112.7355)
			rgb255=(54.927,136.476,109.6245)
			rgb255=(56.125499999999995,137.92950000000002,106.51350000000001)
			rgb255=(57.324,139.4085,103.377)
			rgb255=(58.5225,140.862,100.266)
			rgb255=(59.721,142.341,97.12950000000001)
			rgb255=(60.945,143.7945,94.0185)
			rgb255=(62.1435,145.27349999999998,90.882)
			rgb255=(63.342000000000006,146.7525,87.771)
			rgb255=(64.5405,148.20600000000002,84.66000000000001)
			rgb255=(65.73899999999999,149.685,81.5235)
			rgb255=(66.9375,151.1385,78.4125)
			rgb255=(68.136,152.6175,75.27600000000001)
			rgb255=(69.8955,154.071,72.16499999999999)
			rgb255=(75.5565,155.54999999999998,70.5585)
			rgb255=(81.2175,157.029,72.063)
			rgb255=(86.904,158.48250000000002,73.542)
			rgb255=(92.565,159.9615,75.021)
			rgb255=(98.226,161.415,76.5)
			rgb255=(103.887,162.894,78.00450000000001)
			rgb255=(109.57350000000001,164.06699999999998,79.48349999999999)
			rgb255=(115.23450000000001,165.18900000000002,80.9625)
			rgb255=(120.74249999999999,166.311,82.11)
			rgb255=(124.95,167.4075,82.926)
			rgb255=(129.1575,168.5295,83.742)
			rgb255=(133.365,169.6515,84.5325)
			rgb255=(137.5725,170.77349999999998,85.3485)
			rgb255=(141.78,171.8955,86.16449999999999)
			rgb255=(145.9875,173.01749999999998,86.95500000000001)
			rgb255=(150.195,174.1395,87.771)
			rgb255=(154.4025,175.2615,88.5615)
			rgb255=(158.60999999999999,176.3835,89.3775)
			rgb255=(162.792,177.48,90.1935)
			rgb255=(166.9995,178.602,90.984)
			rgb255=(171.207,179.724,91.8)
			rgb255=(175.41449999999998,180.846,92.59049999999999)
			rgb255=(179.622,181.968,93.40650000000001)
			rgb255=(183.192,182.42700000000002,94.2225)
			rgb255=(184.263,180.2085,95.01299999999999)
			rgb255=(185.334,178.0155,95.82900000000001)
			rgb255=(186.405,175.8225,96.6195)
			rgb255=(187.476,173.60399999999998,97.4355)
			rgb255=(188.5215,171.411,98.2515)
			rgb255=(189.5925,169.218,99.042)
			rgb255=(190.6635,166.9995,99.858)
			rgb255=(191.7345,164.8065,100.6485)
			rgb255=(193.1115,163.2255,104.5755)
			rgb255=(195.9675,164.3985,110.823)
			rgb255=(198.8235,165.75,117.0705)
			rgb255=(201.67950000000002,167.94299999999998,123.318)
			rgb255=(204.5355,170.1615,129.5655)
			rgb255=(207.366,172.38000000000002,135.813)
			rgb255=(210.222,174.573,142.06050000000002)
			rgb255=(213.078,176.7915,148.308)
			rgb255=(215.934,179.469,154.5555)
			rgb255=(218.79,183.141,160.80300000000003)
			rgb255=(221.646,186.66,167.0505)
			rgb255=(224.4765,190.4595,174.1395)
			rgb255=(227.33249999999998,195.45749999999998,181.815)
			rgb255=(230.18849999999998,200.43,189.516)
			rgb255=(233.0445,205.428,197.1915)
			rgb255=(235.90050000000002,210.42600000000002,204.89249999999998)
			rgb255=(238.75650000000002,216.1635,212.568)
			rgb255=(241.6125,222.0285,220.2435)
			rgb255=(244.443,228.7605,227.9445)
			rgb255=(247.299,236.2065,235.62)
			rgb255=(250.155,243.6525,243.32100000000003)
			rgb255=(253.011,250.9965,250.9965)
		},
	colormap={berlin}{
			rgb255=(158.37591,175.99641,254.87428500000001)
			rgb255=(156.100035,175.75313999999997,253.82037)
			rgb255=(153.81651,175.50375,252.765945)
			rgb255=(151.521,175.250535,251.70846)
			rgb255=(149.21707500000002,174.99324,250.64511)
			rgb255=(146.90244,174.73161,249.57691499999999)
			rgb255=(144.57505500000002,174.46233,248.50361999999998)
			rgb255=(142.236705,174.18999,247.42344)
			rgb255=(139.89045000000002,173.90796,246.33408)
			rgb255=(137.52838500000001,173.619045,245.23452)
			rgb255=(135.15867,173.32120500000002,244.12425000000002)
			rgb255=(132.775185,173.01342,243.00020999999998)
			rgb255=(130.38022500000002,172.69365,241.85883)
			rgb255=(127.97506499999999,172.35654,240.70036499999998)
			rgb255=(125.55384,172.00413,239.520735)
			rgb255=(123.12216,171.63412499999998,238.31637)
			rgb255=(120.675945,171.24015,237.086505)
			rgb255=(118.22055,170.82399,235.82553000000001)
			rgb255=(115.752405,170.37876,234.532425)
			rgb255=(113.274315,169.90012499999997,233.20209000000003)
			rgb255=(110.7822,169.38910499999997,231.82968000000002)
			rgb255=(108.284475,168.84060000000002,230.41443)
			rgb255=(105.77859000000001,168.246705,228.950475)
			rgb255=(103.268625,167.608695,227.43577499999998)
			rgb255=(100.759935,166.917645,225.86625)
			rgb255=(98.25048000000001,166.17686999999998,224.23884)
			rgb255=(95.750715,165.37668,222.553035)
			rgb255=(93.26421,164.51682,220.805265)
			rgb255=(90.795045,163.59576,218.994255)
			rgb255=(88.345515,162.61146000000002,217.120005)
			rgb255=(85.93041,161.56136999999998,215.18302500000001)
			rgb255=(83.54871,160.443195,213.184335)
			rgb255=(81.214185,159.260505,211.12393500000002)
			rgb255=(78.932445,158.012535,209.00514)
			rgb255=(76.69992,156.696735,206.83254000000002)
			rgb255=(74.538795,155.324325,204.606645)
			rgb255=(72.44499,153.888675,202.33485)
			rgb255=(70.43227499999999,152.39667,200.01843)
			rgb255=(68.49172499999999,150.8529,197.661465)
			rgb255=(66.63354,149.26042500000003,195.2739)
			rgb255=(64.86384,147.62153999999998,192.85548)
			rgb255=(63.177015,145.94364,190.413345)
			rgb255=(61.57332,144.22698,187.95183)
			rgb255=(60.056325,142.48074,185.474505)
			rgb255=(58.60971,140.70951,182.988)
			rgb255=(57.248265,138.91125,180.490275)
			rgb255=(55.968675000000005,137.09514,177.98949)
			rgb255=(54.74697,135.26041500000002,175.48845)
			rgb255=(53.59386,133.414215,172.986135)
			rgb255=(52.501695,131.55373500000002,170.48739)
			rgb255=(51.451605,129.69249,167.990685)
			rgb255=(50.45889,127.82078999999999,165.50265)
			rgb255=(49.51386,125.94526499999999,163.01818500000002)
			rgb255=(48.59178,124.067955,160.54086)
			rgb255=(47.71356,122.19115500000001,158.07398999999998)
			rgb255=(46.85676,120.31563000000001,155.611455)
			rgb255=(46.027499999999996,118.44087,153.15861)
			rgb255=(45.228075,116.564835,150.714435)
			rgb255=(44.43732,114.69594000000001,148.278165)
			rgb255=(43.662119999999994,112.83086999999999,145.85133)
			rgb255=(42.90171,110.96886,143.43393)
			rgb255=(42.14946,109.11041999999999,141.020355)
			rgb255=(41.421945,107.25504,138.618765)
			rgb255=(40.683975,105.40935,136.22355)
			rgb255=(39.968444999999996,103.56748499999999,133.83828)
			rgb255=(39.245774999999995,101.72766,131.45862)
			rgb255=(38.53611,99.89803500000001,129.08865)
			rgb255=(37.828230000000005,98.070705,126.72786)
			rgb255=(37.138455,96.25281000000001,124.376505)
			rgb255=(36.434145,94.43746499999999,122.02872)
			rgb255=(35.73519,92.630535,119.694195)
			rgb255=(35.05383,90.83202,117.360945)
			rgb255=(34.368135,89.03630999999999,115.042485)
			rgb255=(33.680145,87.24825,112.731675)
			rgb255=(32.997254999999996,85.469115,110.42571)
			rgb255=(32.317425,83.68972500000001,108.12918)
			rgb255=(31.642950000000003,81.921045,105.84310500000001)
			rgb255=(30.97128,80.158485,103.56672)
			rgb255=(30.319245000000002,78.4023,101.29467)
			rgb255=(29.66058,76.65504,99.03588)
			rgb255=(29.001405,74.91415500000001,96.78423)
			rgb255=(28.352684999999997,73.1799,94.54074)
			rgb255=(27.696315000000002,71.455335,92.31102)
			rgb255=(27.070545,69.74173499999999,90.0864)
			rgb255=(26.43585,68.02788000000001,87.87147)
			rgb255=(25.801665,66.32754,85.66775999999999)
			rgb255=(25.18788,64.634085,83.47221)
			rgb255=(24.568485,62.94675,81.289665)
			rgb255=(23.985045,61.26732,79.11808500000001)
			rgb255=(23.405939999999998,59.600384999999996,76.94829)
			rgb255=(22.82403,57.947475,74.79609)
			rgb255=(22.271955000000002,56.304,72.65307)
			rgb255=(21.711209999999998,54.6618,70.52687999999999)
			rgb255=(21.182595000000003,53.035155,68.403495)
			rgb255=(20.67999,51.415905,66.29796)
			rgb255=(20.178150000000002,49.817055,64.204155)
			rgb255=(19.707929999999998,48.22968,62.115195)
			rgb255=(19.270605,46.650465,60.05301)
			rgb255=(18.868215,45.092925,57.99567)
			rgb255=(18.46455,43.550174999999996,55.955414999999995)
			rgb255=(18.116474999999998,42.022725,53.9325)
			rgb255=(17.790585,40.519754999999996,51.92514)
			rgb255=(17.49759,39.025200000000005,49.933589999999995)
			rgb255=(17.2278,37.566345,47.97162)
			rgb255=(16.999575,36.126104999999995,46.01526)
			rgb255=(16.810364999999997,34.69938,44.093835)
			rgb255=(16.661444999999997,33.327225,42.19179)
			rgb255=(16.552305,31.963994999999997,40.333095)
			rgb255=(16.48218,30.63366,38.49123)
			rgb255=(16.451835000000003,29.342850000000002,36.691694999999996)
			rgb255=(16.46127,28.10661,34.924034999999996)
			rgb255=(16.510995,26.883885000000003,33.208650000000006)
			rgb255=(16.6005,25.716495,31.510095)
			rgb255=(16.672665,24.599595,29.878349999999998)
			rgb255=(16.721369999999997,23.54619,28.30704)
			rgb255=(16.80246,22.491255,26.770410000000002)
			rgb255=(16.92894,21.45417,25.31844)
			rgb255=(17.11254,20.413005,23.926395)
			rgb255=(17.389215,19.405245,22.559849999999997)
			rgb255=(17.7786,18.432165,21.171375)
			rgb255=(18.267944999999997,17.506770000000003,19.77372)
			rgb255=(18.86439,16.58979,18.38805)
			rgb255=(19.53198,15.722534999999999,16.996005)
			rgb255=(20.307434999999998,14.93025,15.588915)
			rgb255=(21.155565,14.19483,14.214975)
			rgb255=(22.066935,13.514235000000001,12.83568)
			rgb255=(23.030325,12.928245,11.485199999999999)
			rgb255=(24.0363,12.432015,10.142115)
			rgb255=(25.071345,11.995455,8.844165)
			rgb255=(26.12679,11.63412,7.66887)
			rgb255=(27.216659999999997,11.399775,6.63306)
			rgb255=(28.30143,11.21286,5.706645)
			rgb255=(29.387475,11.11698,4.88325)
			rgb255=(30.484485,11.109585000000001,4.156245)
			rgb255=(31.57206,11.184555,3.5182350000000002)
			rgb255=(32.666775,11.337045,2.9549399999999997)
			rgb255=(33.740325,11.533394999999999,2.430405)
			rgb255=(34.795004999999996,11.77182,2.013225)
			rgb255=(35.861925,12.08037,1.65801)
			rgb255=(36.945420000000006,12.40167,1.358385)
			rgb255=(38.05365,12.708179999999999,1.10823)
			rgb255=(39.189674999999994,13.004235,0.901935)
			rgb255=(40.368795,13.29315,0.73491)
			rgb255=(41.56857,13.570590000000001,0.602565)
			rgb255=(42.791805,13.831199999999999,0.5005649999999999)
			rgb255=(44.04768,14.068859999999999,0.425595)
			rgb255=(45.339254999999994,14.28459,0.374595)
			rgb255=(46.630065,14.4891,0.3417)
			rgb255=(47.95479,14.68137,0.32181000000000004)
			rgb255=(49.274415,14.921069999999999,0.31263)
			rgb255=(50.608065,15.18525,0.312885)
			rgb255=(51.96339,15.427755,0.32130000000000003)
			rgb255=(53.31846,15.67893,0.33711)
			rgb255=(54.68985,15.991050000000001,0.36006)
			rgb255=(56.073735,16.274865000000002,0.389895)
			rgb255=(57.462975,16.581885,0.42712500000000003)
			rgb255=(58.86828,16.905735,0.472515)
			rgb255=(60.28761,17.249475,0.52734)
			rgb255=(61.71408,17.61846,0.5928749999999999)
			rgb255=(63.158655,17.966790000000003,0.67116)
			rgb255=(64.61445,18.35643,0.76449)
			rgb255=(66.08988000000001,18.7782,0.8759250000000001)
			rgb255=(67.574235,19.185435,1.0085250000000001)
			rgb255=(69.08817,19.626075,1.165605)
			rgb255=(70.61664,20.099610000000002,1.351755)
			rgb255=(72.169335,20.608845,1.5710549999999999)
			rgb255=(73.74498,21.134145,1.828605)
			rgb255=(75.34383,21.694125,2.128995)
			rgb255=(76.97379000000001,22.3023,2.48013)
			rgb255=(78.63868500000001,22.92756,2.9210249999999998)
			rgb255=(80.33112000000001,23.59515,3.3976200000000003)
			rgb255=(82.06053,24.32496,3.930315)
			rgb255=(83.82818999999999,25.09098,4.5339)
			rgb255=(85.63027500000001,25.902900000000002,5.214494999999999)
			rgb255=(87.47418,26.769135,5.9772)
			rgb255=(89.35531499999999,27.7032,6.826605)
			rgb255=(91.27648500000001,28.70382,7.76628)
			rgb255=(93.235395,29.74779,8.815605)
			rgb255=(95.23485000000001,30.847604999999998,9.974324999999999)
			rgb255=(97.273065,32.02953,11.141715000000001)
			rgb255=(99.343665,33.266535,12.360105)
			rgb255=(101.450475,34.54587,13.54968)
			rgb255=(103.5861,35.902725000000004,14.75124)
			rgb255=(105.74595000000001,37.29987,15.992325000000001)
			rgb255=(107.923395,38.754645000000004,17.259674999999998)
			rgb255=(110.118435,40.26603,18.62622)
			rgb255=(112.31322,41.82714,20.048099999999998)
			rgb255=(114.516675,43.418595,21.58422)
			rgb255=(116.71452,45.04983,23.171595)
			rgb255=(118.91007,46.719314999999995,24.820425)
			rgb255=(121.09949999999999,48.42144,26.53632)
			rgb255=(123.27210000000001,50.152635,28.314944999999998)
			rgb255=(125.43705,51.89658,30.138450000000002)
			rgb255=(127.58211000000001,53.660415,32.002755)
			rgb255=(129.71595,55.443375,33.910664999999995)
			rgb255=(131.829135,57.23016,35.858865)
			rgb255=(133.92523500000001,59.043465,37.828995)
			rgb255=(136.003995,60.84912,39.846555)
			rgb255=(138.0672,62.667525,41.879414999999995)
			rgb255=(140.117655,64.495365,43.927575)
			rgb255=(142.15281000000002,66.323205,46.002765000000004)
			rgb255=(144.17139,68.150025,48.1032)
			rgb255=(146.18436,69.987555,50.215619999999994)
			rgb255=(148.18356,71.826615,52.335435000000004)
			rgb255=(150.1746,73.66797,54.474375)
			rgb255=(152.16258000000002,75.50907,56.62377)
			rgb255=(154.14342,77.352975,58.784895000000006)
			rgb255=(156.11814,79.207335,60.958259999999996)
			rgb255=(158.09387999999998,81.056085,63.14259)
			rgb255=(160.06554,82.90866,65.328195)
			rgb255=(162.03669,84.772965,67.527825)
			rgb255=(164.009115,86.63497500000001,69.73995000000001)
			rgb255=(165.983835,88.50131999999999,71.955645)
			rgb255=(167.95855500000002,90.37072500000001,74.176185)
			rgb255=(169.936845,92.246505,76.40820000000001)
			rgb255=(171.91947,94.12585499999999,78.645825)
			rgb255=(173.903625,96.01209,80.890845)
			rgb255=(175.894665,97.89959999999999,83.14096500000001)
			rgb255=(177.88698,99.79526999999999,85.40689499999999)
			rgb255=(179.88567,101.69247,87.66798)
			rgb255=(181.88844,103.599105,89.94359999999999)
			rgb255=(183.89529,105.50803499999999,92.22381)
			rgb255=(185.90877,107.42104499999999,94.50759000000001)
			rgb255=(187.92684,109.34298,96.80208)
			rgb255=(189.9495,111.26899499999999,99.102945)
			rgb255=(191.97700500000002,113.200365,101.4084)
			rgb255=(194.011905,115.13556000000001,103.72048500000001)
			rgb255=(196.049355,117.07661999999999,106.03869)
			rgb255=(198.09521999999998,119.02278,108.36531000000001)
			rgb255=(200.144145,120.973785,110.69345999999999)
			rgb255=(202.198425,122.9304,113.033595)
			rgb255=(204.257295,124.889565,115.378575)
			rgb255=(206.32305,126.85893,117.72636000000001)
			rgb255=(208.39161000000001,128.82778499999998,120.08205)
			rgb255=(210.46629000000001,130.80531000000002,122.44335)
			rgb255=(212.546835,132.78665999999998,124.808475)
			rgb255=(214.63146,134.774385,127.184565)
			rgb255=(216.720675,136.765425,129.56448)
			rgb255=(218.81346,138.761055,131.94924)
			rgb255=(220.91262,140.763315,134.340375)
			rgb255=(223.01484,142.769145,136.73559)
			rgb255=(225.121395,144.779055,139.13896499999998)
			rgb255=(227.23305,146.79585,141.54591)
			rgb255=(229.348785,148.814175,143.96025)
			rgb255=(231.46758,150.84015,146.37969)
			rgb255=(233.587905,152.87046,148.80576)
			rgb255=(235.71384,154.905615,151.239225)
			rgb255=(237.84207,156.9423,153.67549499999998)
			rgb255=(239.97438,158.988165,156.118395)
			rgb255=(242.10898500000002,161.03556,158.56716)
			rgb255=(244.24716,163.088565,161.024085)
			rgb255=(246.38584500000002,165.14514,163.48381500000002)
			rgb255=(248.527845,167.20809,165.95196)
			rgb255=(250.67264999999998,169.274865,168.42418500000002)
			rgb255=(252.81924,171.34444499999998,170.90508)
			rgb255=(254.967615,173.41836,173.38725000000002)
		}
}

\pgfdeclarelayer{background}% determine background layer
\pgfdeclarelayer{foreground}% determine foreground layer
\pgfsetlayers{background,main,foreground}% order of layers

\usepgfplotslibrary{external,fillbetween}
\RequirePackage{shellesc}

\usepackage{xcolor}
\definecolor{UKLred} {RGB}{207, 25,  59}
\definecolor{UKLblue}{RGB}{ 47, 63, 157}
\definecolor{NYUpurple} {RGB}{88,15,139}
\definecolor{Pastrami} {RGB}{229,85,79}
\definecolor{TheLake}{RGB}{72,159,223}
\definecolor{EastRiver}{RGB}{0,115,152}
\definecolor{SpicyMustard}{RGB}{203,160,82}
\definecolor{CentralPark}{RGB}{0,108,91}
\definecolor{ProspectPark}{RGB}{64,192,172}
\definecolor{turquois}{rgb}{0,0.75,0.75}%

\usepackage{colortbl}
\usepackage{upgreek}
\usepackage{xr}
\usepackage{nameref}
\usepackage{currfile}

\makeatletter
\newcommand*{\addFileDependency}[1]{
	\typeout{(#1)}
	\@addtofilelist{#1}
	\IfFileExists{#1}{}{\typeout{No file #1.}}
}
\makeatother

\begin{document}

\title{Magnetization transfer explains most of the $T_1$ variability in the MRI literature}

\author[1,2]{Jakob Assl\"ander*}{\orcid{0000-0003-2288-038X}}
\author[1,2]{Sebastian Flassbeck}{\orcid{0000-0003-0865-9021}}
\authormark{Jakob Assl\"ander}

\address[1]{\orgdiv{Center for Biomedical Imaging, Dept. of Radiology}, \orgname{NYU School of Medicine}, \orgaddress{\state{NY}, \country{USA}}}
\address[2]{\orgdiv{Center for Advanced Imaging Innovation and Research (CAI\textsuperscript{2}R), Dept. of Radiology}, \orgname{NYU School of Medicine}, \orgaddress{\state{NY}, \country{USA}}}

\corres{*Jakob Assl\"ander, Center for Biomedical Imaging, NYU School of Medicine, 227 E 30 Street, New York, NY 10016, USA.\\ \email{jakob.asslaender@nyumc.org}}

\finfo{This work was supported by \fundingAgency{NIH/NINDS} grant \fundingNumber{R01~NS131948} and \fundingAgency{NIH/NIBIB} grant \fundingNumber{P41~EB017183}.}

\abstract[Abstract]{
    \textbf{Purpose:}
    To identify the predominant source of the $T_1$ variability described in the literature, which ranges from 0.6--1.1\,s for brain white matter at 3\,T.

    \textbf{Methods:}
    25 $T_1$-mapping methods from the literature were simulated with a mono-exponential and various magnetization-transfer (MT) models, each followed by mono-exponential fitting. A single set of model parameters was assumed for the simulation of all methods, and these parameters were estimated by fitting the simulation-based to the corresponding literature $T_1$ values of white matter at 3\,T. We acquired in vivo data with a quantitative magnetization transfer and three $T_1$-mapping techniques. The former was used to synthesize MR images that correspond to the three $T_1$-mapping methods. A mono-exponential model was fitted to the experimental and corresponding synthesized MR images.

    \textbf{Results:}
    Mono-exponential simulations suggest good inter-method reproducibility and fail to explain the highly variable $T_1$ estimates in the literature. In contrast, MT simulations suggest that a mono-exponential fit results in a variable $T_1$ and explain up to 62\% of the literature's variability. In our own in vivo experiments, MT explains 70\% of the observed variability.

    \textbf{Conclusion:}
    The results suggest that a mono-exponential model does not adequately describe longitudinal relaxation in biological tissue. Therefore, $T_1$ in biological tissue should be considered only a \textit{semi-quantitative} metric that is inherently contingent upon the imaging methodology; and comparisons between different $T_1$-mapping methods and the use of simplistic spin systems---such as doped-water phantoms---for validation should be viewed with caution.
}

\maketitle
\footnotetext{Word Count: 2920}

\section{Introduction}
The Bloch equations \cite{Bloch1946} are the bedrock for our understanding of magnetic resonance imaging (MRI).
They are governed by two time constants, $T_1$ and $T_2$, which characterize the relaxation of longitudinal and transverse magnetization, respectively.
Clinical MRI protocols rely on spin relaxation in the form of $T_1$- and $T_2$-\textit{weighted} images.
Quantification of these parameters, which has been desired since MRI's inception, promises a more objective assessment of the biochemical environment of tissue, and the hypothesis that $T_1$ and $T_2$ are quantitative biomarkers motivates their use in large, multi-center studies and artificial intelligence.
However, the widespread adoption of quantitative relaxometry has been hampered by long scan times and considerable variability in parameter estimates, particularly for $T_1$ where the range is 0.6--1.1\,s for brain white matter at 3\,T.\cite{Stikov.2015,Ou.2008,ZavalaBojorquez2016,Teixeira.2019}
While scan times have been progressively reduced, \cite{Ma2013,Cloos.2016,Christodoulou.2018,Assländer.2019} $T_1$ variability remains a key challenge and decades of research have failed to provide a consensus $T_1$ mapping method.

Numerous explanations for this variability have been hypothesized, including inhomogeneities of the radio frequency (RF) field ($B_1^+$), \cite{Stikov.2015} incomplete spoiling, \cite{Yarnykh.2010,Stikov.2015} and magnetization transfer (MT). \cite{Ou.2008,Mossahebi.2014,Gelderen2016,Rioux.2016,Teixeira.2019,Manning2021}
The present paper identifies MT, i.e., the interaction between spins associated with liquids and macromolecules, \cite{Wolff1989,Henkelman1993} as the dominant cause, which has profound implications for our understanding of spin relaxation.
While mono-exponential relaxation, which is ingrained in the Bloch equations, has a theoretical underpinning for pure liquids, \cite{Bloembergen1948} it does not accurately characterize the spin dynamics in biological tissues, resulting in considerable dependency of $T_1$ estimates on the imaging method.

Previous studies analyzed MT in individual $T_1$-mapping methods. \cite{Ou.2008,Teixeira.2019,Reynolds.2023} Here, we analyze a representative set of the prevalent methods in the literature and demonstrate that MT explains 62\% of the reported $T_1$ variability.
This finding is confirmed with in vivo experiments in which MT explains 70\% of the observed variability.
The best results are achieved when incorporating two recent advances in our understanding of MT: the discovery that the $T_1$ of different spin pools differ substantially \cite{Helms2009,Gelderen2016,Assländer.2024o} and that RF pulses rotate the magnetization of the macromolecular pool rather than saturate it, as described by the generalized Bloch model. \cite{Assländer.2022u6f}
This result suggests that $T_1$ in biological tissue should be considered only a \textit{semi-quantitative} metric.
The following sections discuss implications for the interpretation of past $T_1$ mapping studies and provide suggestions for future directions, including measures for improved inter-study comparability and avenues toward developing methods for fully quantitative biomarkers.

\section{Methods}
\subsection{Fit to literature values}
This study focuses on $T_1$ mapping of brain white matter at 3\,T, for which 25 methods were selected from the literature, including different implementations of
inversion-recovery, \cite{Stanisz.2005,Stikov.2015,Preibisch.2009ng,Shin.2009,Lu.2005,Reynolds.2023,Wright.2008}
Look-Locker, \cite{Stikov.2015,Shin.2009}
saturation-recovery, \cite{Reynolds.2023}
variable flip angle, \cite{Stikov.2015,Cheng.2006,Preibisch.2009ng,Teixeira.2019,Preibisch.2009,Chavez.2012}
MP-RAGE, \cite{Wright.2008}
and MP\textsubscript{2}RAGE. \cite{Marques.2010}
Different implementations of the same techniques vary in shape, amplitude, and timing of RF pulses.
The signal of each method was simulated with various MT models \cite{Graham1997,Assländer.2022u6f} with an emphasis on capturing the RF scheme adequately, while neglecting imaging gradients and assuming complete spoiling as well as homogeneous $B_0$ and $B_1^+$ fields.
Sequence details such as timing, RF pulse shapes, and amplitudes were extracted from the publications and complemented with information kindly provided by authors in private communications.
Missing information was filled heuristically and the source of information for each sequence detail is denoted in the publicly available simulation code (cf. Appendix \ref{sec:data_availability}).
Sequence-specific $T_1$ values were estimated from the simulated data of each pulse sequence with the fitting procedures described in the respective publication.

The sequence-specific $T_1$ estimation is nested in a global---i.e., considering all $T_1$-mapping methods jointly---least-squares fitting routine that estimates a single set of MT and relaxation parameters used to simulate the raw MR signal of each pulse sequence.
The fitting routine chooses the global set of parameters that best explains the $T_1$ variability in the literature with sequence-specific $T_1$ estimates based on these simulations.
Both transversal relaxation times and the exchange rate were fixed in the fit to ensure stability (Tab.~\ref{tab:qMT_param}).

This procedure was repeated with 3 models: a mono-exponential model, Graham's spectral MT model, \cite{Graham1997,Assländer.2022u6f}, and the generalized Bloch MT model. \cite{Assländer.2022u6f}
The source code also contains simulations with Sled's MT model. \cite{Sled2001} As the results are similar to Graham's model, we do not discuss them separately in this paper.

The MT effect was simulated during all RF pulses, including inversion, excitation, and refocussing pulses. Since many pulses are on-resonant, Graham's spectral model was used rather than the more common Graham's single-frequency approximation. As described in Ref.~\citen{Assländer.2022u6f}, the former is an intermediate step in Graham's original publication, \cite{Graham1997} which takes the integral over the line shape, multiplied by the RF pulse's power spectral density. This approach integrates over the singularity of the super-Lorentzian line shape, which is well-defined and numerically stable.

Graham's and Sled's models capture MT as an exponential decay of the semi-solid pool's longitudinal magnetization without explicitly modeling its transversal magnetization. In contrast, the generalized Bloch model builds on the original Bloch model by capturing RF pulses as rotations and modeling the effect of $T_2^s$ explicitly as the relaxation of the pool's transversal magnetization.
Solving its Green's function generalizes the Bloch equations to non-exponential decays or, equivalently, non-Lorentzian lineshapes, as common for semi-solid spin pools.

The generalized Bloch simulations were performed twice, once with the commonly-used constraint $T_1^s=T_1^f$, i.e., assuming equal relaxation times for both pools and once without this constraint.
Here, the superscripts denote the semi-solid or macromolecular and free pool, respectively.
Graham's spectral MT model was simulated with an unconstrained $T_1^s$.
The constraints $T_1^s=T_1^f$ and, similarly, $T_1^s=1\,\text{s}$ were introduced in the early days of MT research \cite{Henkelman1993, Morrison1995a, Yarnykh.2002} to overcome ill-conditioned fits with data acquired for respective publications.
However, recent publications suggested that $T_1^f \approx 2\,\text{s}$ and $T_1^s \approx 0.3\,\text{s}$ in white matter at 3\,T. \cite{Helms2009,Gelderen2016,Manning2021,Samsonov2021}
Within MT models, we recently showed that these constraints cause an underestimation of the semi-solid spin pool size and $T_1^f$. \cite{Assländer.2024o}

\subsection{In vivo experiments}
We scanned one healthy participant with a 3\,T Prisma scanner (Siemens, Erlangen, Germany; software version XA-30) equipped with the vendor's 32-channel receive coil. The protocol was approved by our institutional review board.
First, we acquired an MP\textsubscript{2}RAGE \cite{Marques.2010} with the vendor's product sequence and the setting $T_\text{I} \in \{0.92, 2.88\}\,\text{s}$ and $T_\text{R} = 5\,\text{s}$. We used an adiabatic inversion pulse and non-selective excitation pulses with flip angles $\in \{4, 5\}\,\text{deg}$, spaced 7.08\,ms apart. The readout train was 192 pulses long with centric ordering. The resolution was 1\,mm isotropic.

Second, we acquired a variable-flip angle FLASH protocol with a custom pulse sequence with the settings $T_\text{R} = 21\,\text{ms}$ and flip angles $\in \{4, 25\}\,\text{deg}$. We use rectangular RF pulses with a duration of 100\,$\upmu\text{s}$.
Parameter fitting was performed as described in Ref.~\citen{Cheng.2006} and corrected for $B_1^+$ based on an external map acquired with a preconditioning RF pulse technique.\cite{Chung2010}
This protocol was acquired with 1.3\,mm isotropic resolution.

Third, we acquired inversion-recovery data with a custom pulse sequence similar to the approach described in Ref.~\citen{Dortch.2018}. After each readout train, the magnetization is saturated with 5 90\,deg pulses paired with crushers. Thereafter, the longitudinal magnetization recovers for $T_\text{D} \in \{684, 4171, 2730, 10\}\,\text{ms}$ before it is inverted by a rectangular inversion RF pulse with a duration of 1\,ms. Thereafter, it recovers for $T_\text{I} \in \{15, 15, 278, 1007\}\,\text{ms}$, followed by a RARE readout.\cite{Hennig1986}
This protocol was acquired with 1.3\,mm isotropic resolution.
We fitted a mono-exponential model to this data, where the fitted parameters are $T_1$, an inversion efficiency that captures $B_1^+$ inhomogeneities, and a scaling factor.\cite{Gochberg.2003}
The fit assumes that the magnetization of both pools is destroyed by the saturation pulses.

Last, we performed a hybrid-state\cite{Assländer.2019} quantitative MT (qMT) measurement with a custom pulse sequence, similar to the one described in Ref.~\citen{Assländer.2024o}. The RF pulse train consists of 3 patterns, each 1142 rectangular RF pulses long. Each pattern starts with a 1\,ms non-selective inversion pulse, followed by RF pulses with a variable flip angle and pulse duration. The first pattern is repeated 60 times, followed by 60 repetitions of the second and third pattern. All flip angles and pulse durations were jointly optimized for a minimal Cram\'er-Rao bound in the qMT parameters, weighted by the squared parameters and averaged.\cite{Assländer.2024}
The RF pulse trains were embedded in a pulse sequence with balanced gradient moments. The $T_\text{R}$ was set to 3.5\,ms, resulting in an overall scan time of 12\,minutes.
We used radial koosh-ball k-space readout with a 2D golden means pattern,\cite{Chan.2009} reordered for reduced eddy-current artifacts.\cite{Flassbeck.2023ilg}
Image reconstruction was performed in a Cram\'er-Rao bound-optimized subspace,\cite{Mao.20245s} followed by parameter fitting with a neural network that was trained to have a reduced bias.\cite{Zhang.2022,Mao.202466}
The fitted model incorporates the generalized Bloch model and does not constrain $T_1^s$.

With the qMT parameter maps and the above-described knowledge of the $T_1$-mapping pulse sequences, we synthesized raw images corresponding to each $T_1$-mapping method with MT simulations and performed identical mono-exponential fits on the synthetic and measured images.
We analyzed the $T_1$ estimates in the white matter of a single transversal slice, based on a freeSurfer segmentation.\cite{Fischl.2004kr7}

\section{Results}
\subsection{Fit to literature values}
\begin{figure}[t!]
    \centering
    \begin{tikzpicture}[scale = 1]
    \sffamily\fontsize{7.5}{10}\selectfont
    \begin{axis}[
            width=\textwidth*0.2,
            height=\textwidth*0.2,
            scale only axis,
            xmin = 0.585,
            xmax = 1.12,
            ymin = 0.585,
            ymax = 1.12,
            xticklabel=\empty,
            ylabel={$T_1^\text{literature}~(\text{s})$},
            ylabel style = {yshift = -0.15cm},
            name=monoexp,
            title={mono-exponential model \\ \;},
            title style = {align = center, yshift=-0.123cm},
        ]

        \addplot[
            visualization depends on={\thisrow{seq_marker}\as\seqmarker},
            point meta=\thisrow{seq_marker},
            scatter/classes={
                    I={mark=*,NYUpurple},
                    L={mark=square*,TheLake},
                    v={mark=diamond*,Pastrami},
                    S={mark=+,ultra thick,SpicyMustard},
                    M={mark=triangle*,black}
                },
            scatter, only marks,
            scatter src=explicit symbolic,
        ]
        table [x=sim_mono_exp, y=meas, meta=seq_marker]{Figures/T1_scatter/T1.txt};
        % \node[anchor=center] at (axis cs: 0.888, 0.735)  {\textbf{(\;)}};

        \draw[black, ultra thick, ->] (axis cs: 0.888 - 0.04, 1.084 - 0.04) -- (axis cs: 0.888 - 0.01, 1.084 - 0.01);

        \draw[black, ultra thick] (axis cs: 0.55, 0.55) -- (axis cs: 1.2, 1.2);
        \node[anchor=north west] at (rel axis cs: 0.025, .975)  {\textbf{a}};

        \setlength\extrarowheight{-6pt}
        % \node[anchor=south east] at (rel axis cs: 1.07, 0)  {\begin{tabular}{l} $\Delta \text{AIC} \stackrel{\text{def}}= 0$ \\ $\Delta \text{BIC} \stackrel{\text{def}}= 0$ \end{tabular}};
    \end{axis}

    \begin{axis}[
            width=\textwidth*0.2,
            height=\textwidth*0.2,
            scale only axis,
            xmin = 0.585,
            xmax = 1.12,
            ymin = 0.585,
            ymax = 1.12,
            xticklabel=\empty,
            yticklabel=\empty,
            legend entries = {IR, LL, vFA, SR, MPR},
            legend pos = north west,
            legend style = {xshift=-1cm, yshift=-0.75cm},
            legend columns=2,
            at=(monoexp.east),
            anchor=west,
            xshift = 0.2cm,
            name=Graham,
            title={Graham's MT model \\ with an unconstrained $T_1^s$},
            title style = {align = center, yshift=-0.2cm},
        ]

        \addplot[
            visualization depends on={\thisrow{seq_marker}\as\seqmarker},
            point meta=\thisrow{seq_marker},
            scatter/classes={
                    I={mark=*,NYUpurple},
                    L={mark=square*,TheLake},
                    v={mark=diamond*,Pastrami},
                    S={mark=+,ultra thick,SpicyMustard},
                    M={mark=triangle*,black}
                },
            scatter, only marks,
            scatter src=explicit symbolic,
        ]
        table [x=sim_unconstr_Graham, y=meas, meta=seq_marker]{Figures/T1_scatter/T1.txt};
        % \node[anchor=center] at (axis cs: 0.978, 0.735)  {\textbf{(\;)}};

        \draw[black, ultra thick, ->] (axis cs: 0.877 - 0.04, 1.084 - 0.04) -- (axis cs: 0.877 - 0.01, 1.084 - 0.01);

        \draw[black, ultra thick] (axis cs: 0.55, 0.55) -- (axis cs: 1.2, 1.2);
        \node[anchor=north west] at (rel axis cs: 0.025, .975)  {\textbf{b}};

        \setlength\extrarowheight{-6pt}
        % \node[anchor=south east] at (rel axis cs: 1.07, 0)  {\begin{tabular}{l} $\Delta \text{AIC} = -18.6$ \\ $\Delta \text{BIC} = -16.1$ \end{tabular}};
    \end{axis}

    \begin{axis}[
            width=\textwidth*0.2,
            height=\textwidth*0.2,
            scale only axis,
            xmin = 0.585,
            xmax = 1.12 - 0.001,
            ymin = 0.585,
            ymax = 1.12,
            xlabel={$T_1^\text{simulation}~(\text{s})$},
            ylabel={$T_1^\text{literature}~(\text{s})$},
            ylabel style = {yshift = -0.15cm},
            at=(monoexp.south),
            anchor=north,
            yshift = -1.25cm,
            name=constr,
            title={generalized Bloch MT model \\ with $T_1^s = T_1^f$ constraint},
            title style = {align = center, yshift=-0.19cm},
        ]

        \addplot[
            visualization depends on={\thisrow{seq_marker}\as\seqmarker},
            point meta=\thisrow{seq_marker},
            scatter/classes={
                    I={mark=*,NYUpurple},
                    L={mark=square*,TheLake},
                    v={mark=diamond*,Pastrami},
                    S={mark=+,ultra thick,SpicyMustard},
                    M={mark=triangle*,black}
                },
            scatter, only marks,
            scatter src=explicit symbolic,
        ]
        table [x=sim_constr_gBloch, y=meas, meta=seq_marker]{Figures/T1_scatter/T1.txt};
        % \node[anchor=center] at (axis cs: 1.030, 0.735)  {\textbf{(\;)}};

        \draw[black, ultra thick, ->] (axis cs: 0.972 - 0.04, 1.084 - 0.04) -- (axis cs: 0.972 - 0.01, 1.084 - 0.01);

        \draw[black, ultra thick] (axis cs: 0.55, 0.55) -- (axis cs: 1.2, 1.2);
        \node[anchor=north west] at (rel axis cs: 0.025, .975)  {\textbf{c}};

        \setlength\extrarowheight{-6pt}
        % \node[anchor=south east] at (rel axis cs: 1.07, 0)  {\begin{tabular}{l} $\Delta \text{AIC} =  -16.8$ \\ $\Delta \text{BIC} = -15.6$ \end{tabular}};
    \end{axis}

    \begin{axis}[
            width=\textwidth*0.2,
            height=\textwidth*0.2,
            scale only axis,
            xmin = 0.585,
            xmax = 1.12,
            ymin = 0.585,
            ymax = 1.12,
            xlabel={$T_1^\text{simulation}~(\text{s})$},
            yticklabel=\empty,
            at=(Graham.south),
            anchor=north,
            yshift = -1.25cm,
            name=unconstr,
            title={generalized Bloch MT model \\ with an unconstrained $T_1^{s}$},
            title style = {align = center, yshift=-0.18cm},
        ]

        \addplot[
            visualization depends on={\thisrow{seq_marker}\as\seqmarker},
            point meta=\thisrow{seq_marker},
            scatter/classes={
                    I={mark=*,NYUpurple},
                    L={mark=square*,TheLake},
                    v={mark=diamond*,Pastrami},
                    S={mark=+,ultra thick,SpicyMustard},
                    M={mark=triangle*,black}
                },
            scatter, only marks,
            scatter src=explicit symbolic,
        ]
        table [x=sim_unconstr_gBloch, y=meas, meta=seq_marker]{Figures/T1_scatter/T1.txt};
        % \node[anchor=center] at (axis cs: 0.969, 0.735)  {\textbf{(\;)}};

        \draw[black, ultra thick, ->] (axis cs: 0.985 - 0.04, 1.084 - 0.04) -- (axis cs: 0.985 - 0.01, 1.084 - 0.01);

        \draw[black, ultra thick] (axis cs: 0.55, 0.55) -- (axis cs: 1.2, 1.2);
        \node[anchor=north west] at (rel axis cs: 0.025, .975)  {\textbf{d}};

        \setlength\extrarowheight{-6pt}
        % \node[anchor=south east] at (rel axis cs: 1.07, 0)  {\begin{tabular}{l} $\Delta \text{AIC} =  -24.7$ \\ $\Delta \text{BIC} = -22.3$ \end{tabular}};
    \end{axis}

\end{tikzpicture}
    \vspace{-0.25cm}
    \caption{
        Literature $T_1$ estimates based on measured data in comparison to $T_1$ estimates from MT simulations. Overall, 25 $T_1$-mapping methods were simulated, comprising different implementations of inversion-recovery (IR), Look-Locker (LL), variable flip angle (vFA), saturation-recovery (SR), and MP\textsubscript{(2)}RAGE (MPR).
        % The violin plots highlight the reduced variability when simulating the signal with an MT model.
        The arrows highlight an IR method with a very short inversion pulse.
    }
    % \vspace{-0.8cm}
    \label{fig:T1_scatter}
\end{figure}
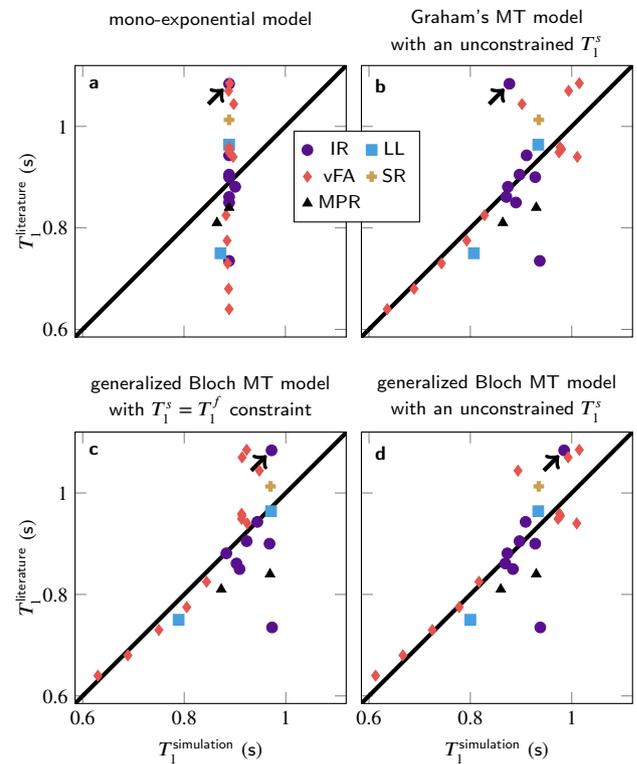

% \cite{Stanisz.2005,Stikov.2015,Lu.2005,Shin.2009,Reynolds.2023,Cheng.2006,Teixeira.2019,Preibisch.2009,Wright.2008,Marques.2010,Preibisch.2009ng,Chavez.2012}
In contrast to the 3\% intra-study coefficients of variation reported for $T_1$, \cite{Gracien.2020} % Vrenken.2006
the inter-study coefficient of variation is 14\%
across the literature analyzed here.
Fig.~\ref{fig:T1_scatter} illustrates this variability by the spread along the y-axis, and compares it to $T_1$ estimates based on signals that were simulated for the respective pulse sequence, followed by mono-exponential fitting as described in the respective publication.
The simulations of all pulse sequences used a global set of model parameters, which was determined with a least-square fitting procedure to best explain the literature $T_1$ values (Tab.~\ref{tab:qMT_param}).

Simulating the signals with a mono-exponential model (Fig.~\ref{fig:T1_scatter}a) results in a small span along the x-axis, indicating inter-study reproducibility within the mono-exponential framework, which matches experimental findings in phantoms containing doped water. \cite{Stikov.2015}
However, the deviations from the identity line indicate that a mono-exponential model fails to explain the inter-study variability observed in tissue.

\begin{table}[t!]
    % \vspace{-0.5cm}
    \newcommand{\g}{\cellcolor[gray]{0.8}}
    \begin{center}
        \begin{tabular}[]{c|c|c|c|c|c|c}
            MT model                & \multicolumn{2}{c|}{Graham's} & \multicolumn{4}{c}{generalized Bloch}                                                                                                     \\
            \midrule
            $T_1^s$ constraint      & \multicolumn{2}{c|}{none}     & \multicolumn{2}{c|}{$T_1^s = T_1^f$}  & \multicolumn{2}{c}{none}                                                                          \\
            \midrule
            study                   & this                          & \cite{Gelderen2016,Stanisz.2005}      & this                     & \cite{Assländer.2024,Stanisz.2005} & this   & \cite{Assländer.2024o} \\
            \midrule
            $m_0^s$                 & 0.19                          & 0.27                                  & 0.13                     & 0.14                               & 0.21   & 0.21                     \\
            $T_1^f$ (s)             & 2.03                          & 2.44                                  & 0.97                     & 1.52                               & 2.06   & 1.84                     \\
            $T_1^s$ (s)             & 0.25                          & 0.25                                  & \g$T_1^f$                &                                    & 0.26   & 0.34                     \\
            $T_2^f$ (ms)            & \g76.9                        & 69                                    & \g76.9                   & 70.1                               & \g76.9 & 76.9                     \\
            $T_2^s$ ($\upmu$s)      & \g12.5                        & 10.0                                  & \g12.5                   &                                    & \g12.5 & 12.5                     \\
            $R_\text{x}$ (s$^{-1}$) & \g13.6                        & 9.0                                   & \g23.0                   & 23.0                               & \g13.6 & 13.6                     \\
            \bottomrule
        \end{tabular}
    \end{center}
    \vspace{-0.25cm}
    \caption{Estimates of MT parameters. The parameters were estimated by fitting MT models to variable literature $T_1$ values (\textit{this}) and are here compared to MT parameters reported in the literature. \cite{Gelderen2016,Stanisz.2005,Assländer.2024,Assländer.2024o} $m_0^s$ denotes the semi-solid spin pool size, the relaxation times $T_{1,2}$ are qualified by the superscripts $^{f,s}$ to identify the free and semi-solid spin pool, respectively, and $R_\text{x}$ is the exchange rate. The gray background highlights parameters that were fixed during the fit.}
    \vspace{-0.2cm}
    \label{tab:qMT_param}
\end{table}

Simulating the signal with various MT models and fitting a mono-exponential model to the simulated data replicates most of the $T_1$ variability (b--d), i.e., the median absolute deviation is reduced by 62\% when comparing the residuals of the generalized Bloch fit without $T_1^s$ constraint to the $T_1$ estimates in the literature.
To provide some context for this result, note that the simulations are based on incomplete knowledge of implementation details, despite many authors kindly providing unpublished information.
Incorrect implementation details can result in outliers, which were not excluded from the least-square fitting of the MT parameters.
Outliers impair the performance of least-square fitting, which intrinsically assumes a Gaussian distribution of residuals.
As the residuals' distribution is unknown, least-squares fitting is used for simplicity, and to ensure a stable fit, literature values were used for the transversal relaxation times and the exchange rate.
The other MT parameters were fitted and aligned well with the literature (Tab.~\ref{tab:qMT_param}).
Removing all constraints further reduces the residuals, at the cost of less plausible MT parameters.

Different MT models capture the $T_1$ variability to slightly different degrees: Graham's spectral model \cite{Graham1997} does not adequately describe the spin dynamics during a 10$\upmu$s inversion-pulse (arrow in Fig.~\ref{fig:T1_scatter}b). This challenge is overcome by the generalized Bloch model \cite{Assländer.2022u6f} (c--d). Further, the commonly-used constraint $T_1^s=T_1^f$ entails larger residuals compared to the recently proposed unconstrained fit (c vs.~d).
The Akaike and Bayesian information criteria (Tab.~\ref{tab:AIC_BIC}) indicate that a fit with the generalized Bloch model and without $T_1^s$ constraint best explains the $T_1$ variability and that the increased number of variables is justified.

\begin{table}[t!]
    \vspace{-0.5cm}
    \begin{center}
        \begin{tabular}[]{c|c|c|c}
            model             & $T_1^s$ constraint & $\Delta$AIC & $\Delta$BIC \\
            \midrule
            mono-exponential  & none               & 0           & 0           \\
            Graham's          & none               & -18.6       & -16.1       \\
            generalized Bloch & $T_1^s = T_1^f$    & -16.8       & -15.6       \\
            generalized Bloch & none               & -24.7       & -22.3       \\
            \bottomrule
        \end{tabular}
    \end{center}
    \vspace{-0.5cm}
    \caption{Akaike (AIC) and Bayesian (BIC) information criteria. The values are relative to the mono-exponential fit ($\Delta$AIC = AIC -- AIC\textsubscript{mono}) and lower values indicate a preferable model. AIC and BIC weigh residuals against the number of model parameters and the results indicate that the generalized Bloch model without $T_1^s$ constraint is preferable despite the penalty for its larger number of parameters.}
    \vspace{-0.25cm}
    \label{tab:AIC_BIC}
\end{table}

\subsection{In vivo experiments}
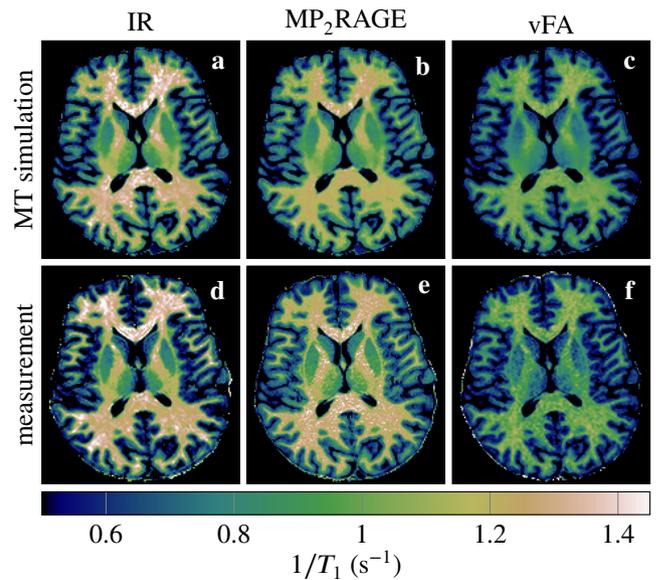
\begin{figure}[bp]
    \centering
    \begin{tikzpicture}[scale = 1, every text node part/.style={align=left}]
    \begin{axis}[%
            width={2.6cm},
            height={2.6cm*165/150},
            axis on top,
            scale only axis,
            xmin=0,
            xmax=1,
            ymin=0,
            ymax=1,
            xtick = \empty,
            ytick = \empty,
            ylabel={MT simulation},
            title={IR},
            title style={yshift = -0.2cm},
            name=IR_sim_R1,
        ]
        \addplot graphics [xmin=0,xmax=1,ymin=0,ymax=1] {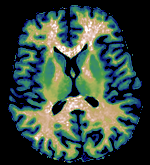};
        \node[fill=black,text=white, opacity=0.75, text opacity=1, anchor = north east] at (rel axis cs:  0.975,0.975) {\textbf{a}};
    \end{axis}

    \begin{axis}[%
            width={2.6cm},
            height={2.6cm*165/150},
            axis on top,
            scale only axis,
            xmin=0,
            xmax=1,
            ymin=0,
            ymax=1,
            xtick = \empty,
            ytick = \empty,
            ylabel={measurement},
            name=IR_meas_R1,
            at=(IR_sim_R1.south),
            anchor=north,
            yshift = -0.1cm,
        ]
        \addplot graphics [xmin=0,xmax=1,ymin=0,ymax=1] {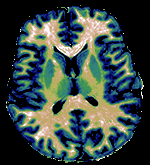};

        \node[fill=black,text=white, opacity=0.75, text opacity=1, anchor = north east] at (rel axis cs:  0.975,0.975) {\textbf{d}};
    \end{axis}

    \begin{axis}[%
            width={2.6cm},
            height={2.6cm*165/150},
            axis on top,
            scale only axis,
            xmin=0,
            xmax=1,
            ymin=0,
            ymax=1,
            title={MP\textsubscript{2}RAGE},
            title style={yshift = -0.25cm},
            xtick = \empty,
            ytick = \empty,
            name=MPR_sim_R1,
            at=(IR_sim_R1.east),
            anchor=west,
            xshift = 0.1cm,
        ]
        \addplot graphics [xmin=0,xmax=1,ymin=0,ymax=1] {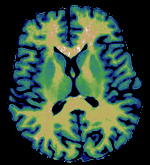};

        \node[fill=black,text=white, opacity=0.75, text opacity=1, anchor = north east] at (rel axis cs:  0.975,0.975) {\textbf{b}};
    \end{axis}

    \begin{axis}[%
            width={2.6cm},
            height={2.6cm*165/150},
            axis on top,
            scale only axis,
            xmin=0,
            xmax=1,
            ymin=0,
            ymax=1,
            xtick = \empty,
            ytick = \empty,
            name=MPR_meas_R1,
            at=(MPR_sim_R1.south),
            anchor=north,
            yshift = -0.1cm,
        ]
        \addplot graphics [xmin=0,xmax=1,ymin=0,ymax=1] {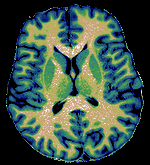};

        \node[fill=black,text=white, opacity=0.75, text opacity=1, anchor = north east] at (rel axis cs:  0.975,0.975) {\textbf{e}};
    \end{axis}

    \begin{axis}[%
            width={2.6cm},
            height={2.6cm*165/150},
            axis on top,
            scale only axis,
            xmin=0,
            xmax=1,
            ymin=0,
            ymax=1,
            xtick = \empty,
            ytick = \empty,
            title={vFA},
            title style={yshift = -0.25cm},
            name=vFA_sim_R1,
            at=(MPR_sim_R1.east),
            anchor=west,
            xshift = 0.1cm,
        ]
        \addplot graphics [xmin=0,xmax=1,ymin=0,ymax=1] {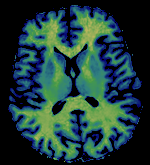};
        \node[fill=black,text=white, opacity=0.75, text opacity=1, anchor = north east] at (rel axis cs:  0.975,0.975) {\textbf{c}};
    \end{axis}

    \begin{axis}[%
            width={2.6cm},
            height={2.6cm*165/150},
            axis on top,
            scale only axis,
            xmin=0,
            xmax=1,
            ymin=0,
            ymax=1,
            xtick = \empty,
            ytick = \empty,
            name=vFA_meas_R1,
            at=(MPR_meas_R1.east),
            anchor=west,
            xshift = 0.1cm,
            colormap name = gist_earth,
            colorbar horizontal,
            point meta min=0.5,
            point meta max=1.45,
            colorbar style={xlabel=$1/T_1~(\text{s}^{-1})$, height=0.3cm, yshift=0.2cm, width=8cm, xshift=-5.4cm, xlabel style = {yshift = 0.15cm}, xticklabel style={/pgf/number format/fixed, /pgf/number format/precision=2}},
        ]
        \addplot graphics [xmin=0,xmax=1,ymin=0,ymax=1] {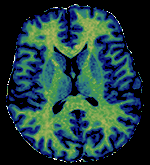};

        \node[fill=black,text=white, opacity=0.75, text opacity=1, anchor = north east] at (rel axis cs:  0.975,0.975) {\textbf{f}};
    \end{axis}

\end{tikzpicture}
    \vspace{-0.8cm}
    \caption{
        In vivo $T_1$ maps based on MR images synthesized with MT simulations (\textbf{a}--\textbf{c}) and measured (\textbf{d}--\textbf{f}). The three columns compare an inversion-recovery (IR), an MP\textsubscript{2}RAGE, and a variable flip angle (vFA) pulse sequence.
    }
    \vspace{-0.2cm}
    \label{fig:T1_InVivo}
\end{figure}

We fitted a mono-exponential model to measured MR data acquired with different pulse sequences. Comparing the resulting $T_1$ maps to one another (Fig.~\ref{fig:T1_InVivo}) confirms the variability observed in the literature in the absence of biological variability.
Performing the same mono-exponential fits on MRI images synthesized with MT simulations---based on the generalized Bloch model and an unconstrained $T_1^s$---results in $T_1$ maps that closely resemble the measured $T_1$ maps. The biggest deviations are observed for the MP\textsubscript{2}RAGE sequence.

\begin{figure}[t!]
    \centering
    \begin{tikzpicture}[scale = 1]
    \sffamily\fontsize{7.5}{10}\selectfont
    \begin{axis}[
        width=\textwidth*0.2,
        height=\textwidth*0.2,
        scale only axis,
        xmin = 0.75,
        ymin = 0.75,
        xmax = 1.05,
        ymax = 1.05,
        xlabel={$T_1^\text{simulation}~(\text{s})$},
        ylabel={$T_1^\text{measurement}~(\text{s})$},
        ylabel style = {yshift = -0.15cm},
        title={generalized Bloch MT model \\ with an unconstrained $T_1^{s}$},
        title style = {align = center, yshift=-0.18cm},
        legend entries = {IR, vFA, MPR},
        legend pos = north west,
        ]

        \addplot[
            visualization depends on={\thisrow{seq_marker}\as\seqmarker},
            point meta=\thisrow{seq_marker},
            scatter/classes={
                    I={mark=*,NYUpurple},
                    v={mark=diamond*,Pastrami},
                    M={mark=triangle*,black}
                },
            scatter, only marks,
            scatter src=explicit symbolic,
        ]
        table [x=sim, y=meas, meta=seq_marker]{Figures/T1_scatter_invivo/T1_invivo.txt};

        \draw[black, ultra thick] (axis cs: 0.55, 0.55) -- (axis cs: 1.2, 1.2);
    \end{axis}
\end{tikzpicture}
    \vspace{-0.25cm}
    \caption{White matter ROI analysis of the $T_1$ maps shown in Fig.~\ref{fig:T1_InVivo}, comparing mono-exponential $T_1$ estimates based on MT-simulated and measured MR images. Here, we compare an inversion-recovery (IR), an MP\textsubscript{2}RAGE (MPR), and a variable flip angle (vFA) pulse sequence.}
    % \vspace{-0.8cm}
    \label{fig:T1_scatter_invivo}
\end{figure}
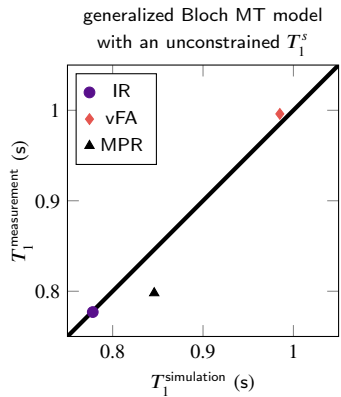

The median $T_1$ of all white matter voxels in a transversal slice (cf. Fig.~\ref{fig:T1_InVivo}) is analyzed in Fig.~\ref{fig:T1_scatter_invivo}. It confirms the agreement between MT simulations and experiments. It also confirms that the MP\textsubscript{2}RAGE exhibits the biggest deviations.
Analyzing the median absolute deviation, we find that MT explains 70\% of the variability observed in these experiments.

\section{Discussion}
Only one year after the discovery of MT, \cite{Wolff1989} Koenig et al.~\cite{Koenig1990} hypothesized an association between MT and $T_1$ relaxation.
Notwithstanding, MT has traditionally been considered a nuisance effect in $T_1$ mapping and, likely due to time constraints of \textit{in vivo} imaging settings, most methods assume a mono-exponential model.
However, recent studies picked up on Koenig's hypothesis and suggested that MT is an integral driver of longitudinal relaxation. \cite{Helms2009,Gelderen2016,Assländer.2024o}
This paper analyzes the variability in mono-exponential $T_1$ estimates and links it to pervasive but variable contributions of MT.

In the absence of RF pulses, e.g., during an inversion-recovery experiment, the two-pool MT model describes bi-exponential relaxation. \cite{Henkelman1993,Gochberg.2003}
Fitting a mono-exponential model to such data elicits a sensitivity of the estimated $T_1$ to the inversion times, explaining the observed variability.
This brings into question the common classification of the inversion-recovery method with mono-exponential fitting as the gold standard for $T_1$ mapping in biological tissue.

RF pulses affect the two spin pools differently due to their vastly different $T_2$ relaxation times (10$\upmu$s vs. 100ms).
As a consequence, the measured signal is sensitive to the shape and amplitude of the RF pulses, as well as the timing of their sequence.
This sensitivity includes inversion-recovery methods and is pronounced for variable flip angle methods, which rely on many RF pulses in rapid succession.

The finding that MT explains most of the $T_1$ variability indicates that the principal cause is an oversimplified model rather than experimental limitations, which positions $T_1$ in biological tissue as a \textit{semi-quantitative} metric, inherently contingent upon the employed imaging methodology.
It questions the comparability of different $T_1$-mapping techniques and suggests that validations conducted in simplistic spin systems, such as doped-water phantoms, might provide only a partial assessment of $T_1$-mapping methods.

It is important to note that different imaging methods do not result in re-scaled versions of the same $T_1$. On the contrary, different methods capture different weightings of the individual relaxation mechanisms
and might have different sensitivities to pathology, making them fundamentally incomparable.
Notably, even small variations in the data acquisition protocol can influence the contributions of different relaxation mechanisms as exemplified by the inversion-recovery method: short inversion times are sensitive to the exchange rate, while long inversion times are mostly sensitive to the spin-pool size $m_0^s$ and the pools' relaxation times $T_1^{f,s}$. \cite{Gochberg.2003,Assländer.2024o}
For most methods, however, the composition of relaxation mechanisms is not intuitively evident and an analysis would require consideration of all sequence aspects as well as the parameter fitting routine, which is beyond the scope of this paper.

One path toward more reproducible $T_1$ mapping would be to design methods in which each data point has a similar sensitivity to the MT parameters.
For inversion-recovery methods, this could be achieved by acquiring data only at inversion times much longer than the fast component, i.e., much longer than 100\,ms. \cite{Gochberg.2003,Assländer.2024o}
For variable flip angle methods, Teixeira et al. \cite{Teixeira.2019} suggested adding off-resonant saturation to each RF pulse such that the macromolecular spin pool is kept constant over variable flip angles.
The resulting relaxation model is mono-exponential with a composition of relaxation mechanisms that depend on the applied RF power.
Teixeira et al. proposed to further qualify the reported $T_1$ values by the applied RF power to identify studies that assess similar compositions.

Phantom validation studies could be improved by replacing doped water with gels that have a semi-solid spin pool. A common choice is cross-linked bovine serum albumin, which is well-described by a super-Lorentzian lineshape. Agar is also a popular choice. However, due to its chemical structure, it is better described by a Gaussian lineshape.\cite{Henkelman1993}

The here-presented MT simulations explain 62\% of the literature's inter-study $T_1$ variability. The residual variability likely has multiple sources.
First, the publicly available sequence details and the information kindly provided by the authors are incomplete and several sequence details are heuristically chosen, as noted in the simulation code. For this reason, MT could explain a larger fraction of the variability with more accurate sequence information, which is supported by better fit (70\%) in our own experiments where all sequence information was available.
Second, further unmodeled effects such as dipolar coupling (cf. next paragraph) are potential contributors.
Third, experimental imperfections, such as incomplete spoiling, inaccurate $B_1^+$ maps, etc., likely contribute to the residual variability.
Last, biological variability likely contributes to the literature's variability, as most of the analyzed values originate from individual participants or small cohorts that are not matched between studies.

In our own experiments, MT explains 70\% of the observed inter-sequence variability. We observe the biggest deviations in the MP\textsubscript{2}RAGE sequence and excluding this sequence increases the explained variability to 94\%.
In our experiments, the MP\textsubscript{2}RAGE is the only pulse sequence that utilizes an adiabatic RF pulse.
MT simulations of such pulses suggest a near-complete saturation of the semi-solid pool ($< 1\%$ of the initial $z$-magnetization remains).
This stands in contrast to the experimental findings of Reynolds et al.\cite{Reynolds.2023} who recently reported that 22--24\% of the magnetization remains after an adiabatic inversion pulse. This discrepancy could explain that the MP\textsubscript{2}RAGE and other sequences with adiabatic pulses are outliers.
To the best of our knowledge, Reynold's findings cannot be explained with a 2-pool MT model and reasonable relaxation times, but dipolar order effects could explain those results. \cite{Morrison1995,Varma2015,Manning2017}
Exchange with the dipolar order is facilitated by off-resonant RF irradiation, like at the beginning and end of an adiabatic pulse's frequency sweep.

It is noted that any model entails simplifications, especially considering the complexity of biological tissue. For example, Manning et al. demonstrated in a post-mortem NMR study that a 4-pool model describes white matter more accurately than a 2-pool model. \cite{Manning2017}
In vivo, scan time and hardware constraints limit the ability to acquire sufficient data for complex models and further research is needed to identify adequate compromises between model complexity and method-dependent bias.

This paper focuses on established $T_1$-mapping methods. More recently, multi-parametric approaches that simultaneously estimate $T_1$ and $T_2$ have gained popularity. For such sequences, model oversimplifications can bleed into the estimates of $T_2$, as demonstrated in Ref.~\citen{Hilbert.2019}.

\section{Conclusion}
This paper provides a comprehensive comparison of established $T_1$-mapping methods and identifies the relaxation model as the principal bottleneck on the road to quantitative biomarkers.
The presented findings suggest that a separation of the individual relaxation mechanisms, as performed in quantitative MT, is necessary to quantify longitudinal relaxation without major dependencies on implementation details.

\appendix
\section{Acknowledgements}
The authors would like to thank Drs.
Stanisz,
Stikhov, Boudreau, Leppert,
Marques,
Malik, Teixeira,
Michal, Reynolds,
van Zijl,
Cheng,
Gowland,
Preibisch, Deichmann,
and Shin
for providing unpublished implementation details of their $T_1$-mapping methods.

The authors thank the ISMRM Reproducible Research Study Group for conducting a code review of the code (version 0.2) supplied in the Data availability statement. The scope of the code review covered only the code’s ease of download, quality of documentation, and ability to run, but did not consider scientific accuracy or code efficiency.

\section{Data availability statement} \label{sec:data_availability}
Code to replicate all results can be found at \url{https://jakobasslaender.github.io/T1variability/v1.0}. This website outlines all simulation code along with the presented results.
The results in the present paper were created with v1.0 of the simulation code.
Contributions of additional $T_1$-mapping methods as well as improvements of the current methods are explicitly welcomed and can be facilitated via GitHub pull-requests. The fitting results are continuously updated with continuous-integration tools.

\bibliography{My_Library.bib}
\end{document}